\documentclass{article}
\usepackage{cite}
\usepackage{amsmath,amssymb,amsfonts}
\usepackage{algorithm}
\usepackage{graphicx}
\usepackage{subcaption}

\usepackage{textcomp}
\usepackage{xcolor}
\usepackage{array}
\usepackage{hyperref}
\usepackage{booktabs}
\usepackage{balance}
\usepackage{algpseudocode}
\usepackage{perpage}
\date{}

\begin{document}
\title{Federated Deep Reinforcement Learning - based Bitrate Adaptation for Dynamic Adaptive Streaming over HTTP}

\author{
Phuong L. Vo, Nghia T. Nguyen, Long Luu, \\
Canh T. Dinh, Nguyen H. Tran, Tuan-Anh Le
\thanks{Phuong L. Vo, Nghia T. Nguyen, Long Luu are with the International University, Vietnam National University, Ho Chi Minh City, Vietnam; Canh T. Dinh, Nguyen H. Tran are with the University of Sydney, Sydney, NSW 2006, Australia; Tuan-Anh Le is with Institute of Engineering and Technology, Thu Dau Mot University, Binh Duong, Vietnam.}
}

\maketitle              

\begin{abstract}
In video streaming over HTTP, the bitrate adaptation selects the quality of video chunks depending on the current network condition. Some previous works have applied deep reinforcement learning (DRL) algorithms to determine the chunk's bitrate from the observed states to maximize the quality-of-experience (QoE). However, to build an intelligent model that can predict in various environments, such as 3G, 4G, Wifi, \textit{etc.}, the states observed from these environments must be sent to a server for training centrally.

In this work, we integrate federated learning (FL) to DRL-based rate adaptation to train a model appropriate for different environments. The clients in the proposed framework train their model locally and only update the weights to the server. The simulations show that our federated DRL-based rate adaptations, called FDRLABR with different DRL algorithms, such as deep Q-learning, advantage actor-critic, and proximal policy optimization, yield better performance than the traditional bitrate adaptation methods in various environments.

\textit{Keywords: bitrate adaptation, deep reinforcement learning, federated learning, dynamic adaptive streaming over HTTP}
\end{abstract}

\section{Introduction}

Dynamic adaptive streaming over HTTP (DASH) is the primary method of video streaming on the Internet today. This standard is widely applied because of its flexibility and scalability. The video is chunked and encoded in different bitrates. Depending on the current network condition, the user's adaptive bitrate (ABR) function chooses an appropriate bitrate to request the video chunks \cite{smooth, video, bola}.

The traditional ABR methods include throughput-based~\cite{smooth} and buffer-based methods~\cite{bola}. The throughput-based method chooses the quality level of the next chunk using the estimated throughput (usually the mean) of the previously downloaded chunks~\cite{smooth}. The buffer-based method determines the quality level of the next chunk based on the current buffer level. BOLA is a well-known buffer-based method that optimizes the Lyapunov function to select a quality level for chunks~\cite{bola}.
Both throughput-based and BOLA are deployed as two main ABR methods in the reference video client Dash.js~\cite{dashjs}. 

The ABR based on tabular Q-learning maximizing the user's QoE is studied in \cite{Claeys14}. The states, including estimated throughput and buffer size, are discretized.
The reward combines three objectives: high chunk quality level, low number of quality switches, and short rebuffering time.
The success of deep Q-learning (DQN), which achieved human-level playing in the Atari game~\cite{dqn}, has inspired the development of various deep reinforcement learning (DRL) algorithms and their applications in many fields.
Recently, there have been several works \cite{pensieve, ddash, liu2018} that applied several DRL algorithms to improve the performance of ABR.
In work \cite{pensieve}, the proposed Pensieve method has applied an asynchronous advantage actor-critic (A3C) algorithm to rate adaptation. The action is the quality level the client will request for the next chunk.
The state combines different observations, such as estimated throughput, next chunk sizes, number of remaining chunks, buffer size, delay, \textit{etc.}
The reward is the utility penalized by quality switching and video rebuffering. Since A3C is a policy-based method, a neural network approximates a state to an action distribution, and the action corresponding to the maximum probability is chosen.
Similar to Pensieve, D-DASH proposed in \cite{ddash} uses DQN algorithm for ABR. The state space, action space, and reward function are similar to the ones in Pensieve. Two network architectures are used in \cite{ddash}, \textit{i.e.}, multilayer perceptron and long-short term memory.
The work \cite{liu2018} has applied several DRL algorithms to ABR, \textit{i.e.}, temporal difference, A3C, DQN, and rainbow DQN.
The DRL-based ABR methods achieve a higher QoE than the traditional methods.

The federated learning (FL) algorithms \cite{McMahan17} in supervised learning emerge as promising algorithms that allow learning distributively.
In each round, several random clients are selected to train in several epochs with their local datasets and send the weights of the trained models to the server. The server then averages the received weights and broadcasts the average weight to all clients.
In FL, data is kept locally without being sent to the server, hence, preserving the user’s privacy.

Inspired by the success of FL, federated deep reinforcement learning (FDRL) is the architecture combining federated learning and reinforcement learning \cite{fdrlsurvey}.
Inherit from the categorization of FL algorithms \cite{flsurvey}, FDRL has two groups of algorithms: horizontal FDRL and vertical FDRL \cite{fdrlsurvey}.
In horizontal FDRL, the clients have aligned the state space and action space. The clients explore different aspects of their environment but are unwilling to share their experiences with a central model. The horizontal FDRL framework helps to increase training samples and improve global model performance. 
In vertical FDRL, clients' training data may have different features.  The state space and the action space of the clients are not aligned. Each client observes part of the global environment and can perform a different set of actions depending on their observed environment.
 
We propose a horizontal FDRL architecture for rate adaptation in which the clients experience different environments like network delay, throughput, buffer thresholds, \textit{etc.}
The contributions in this paper include the following:
\begin{enumerate}
	\item We propose a federated deep reinforcement learning-based bite rate adaptation framework called FDRLABR. The global model can predict action in various environments.
	\item We implement FDRLABR with different DRL algorithms run at clients, including value-based algorithm, \textit{i.e.}, deep Q-learning (DQN), and actor-critic algorithms, \textit{i.e.}, advantage actor-critic (A2C) and proximal policy optimization (PPO)\footnote{\url{https://github.com/toiuuvagiaithuat/FDRLABR}}.
	\item We train and evaluate our proposed algorithms in an event-driven environment. The global model can predict the action for various environment datasets and yields a better QoE than the traditional methods such as BOLA and throughput-based.
\end{enumerate}
To the best of our knowledge, this study is the first work that proposes a distributed DRL framework for bitrate adaptation in DASH.

The remaining paper has the following structure. Section \ref{sec:fdrlabr} describes FDRLABR framework. Section \ref{sec:result} presents the performance of the proposed algorithms, and Section \ref{sec:con} concludes the work.

\section{Federated deep reinforcement learning-based bitrate adaptation (FDRLABR)}
\label{sec:fdrlabr}
\begin{figure}
	\centering
	\includegraphics[width=11cm]{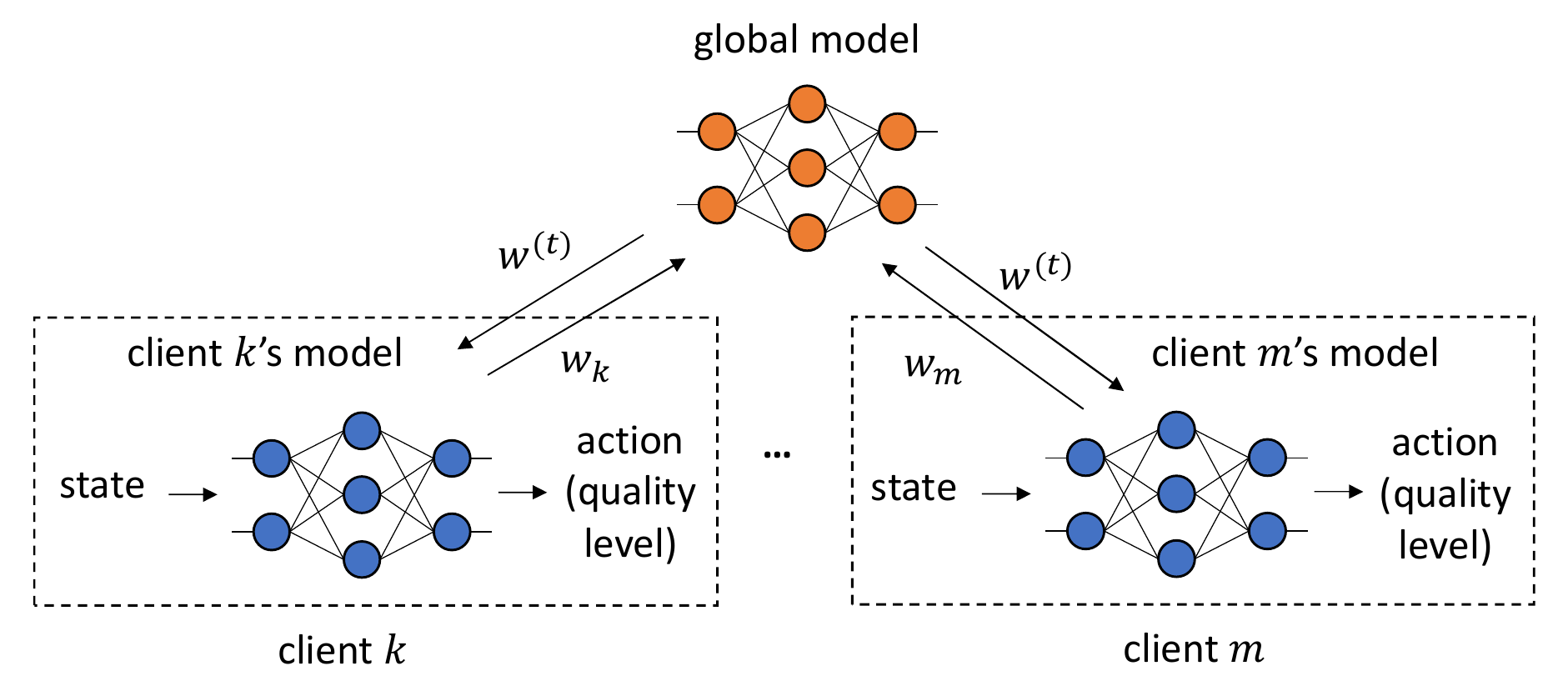}
	\caption{The framework of FDRLABR.}
	\label{fig:framework}	
\end{figure}

Our proposed FDRLABR framework includes a global model at the server implementing federated updates and multiple local models at clients running DRL algorithms.
Fig. \ref{fig:framework} describes the framework of FDRLABR. 
Both global and local models have the same neural network architecture. 
In each global round, the server randomly selects several clients and broadcasts its model's parameters to these clients. Each selected client performs local training for $E$ episodes and sends back the weights of its model weights to the server. The server then averages the received weights and goes to the next round.

\subsection{DRL model for bitrate adaptation at client}
A DRL agent interacts with the environment and learns from experiences.
At each time step, the agent observes state~$s$, chooses action $a$, and receives a reward $r$. The agent aims to maximize the cumulative reward \cite{Sutton}.
Assuming that each time step begins when the client requests a video chunk and an episode is the set of all the chunks of a video. The reward function, states, and actions of the DRL model applied for bitrate adaptation are described as follows.

Corresponding to an observed state, the DRL agent decides which quality level of the chunk will be requested. 
The \textit{action space} includes all the quality levels of the video.
At time step $n$, assuming that the agent takes the action corresponding download chunk $n$ at bitrate $R_n$, the received \textit{reward} is the combination of three QoE metrics: the utility value corresponding to chunk’s quality level, the quality difference between two consecutive chunks, and the rebuffering time in this time step. The following formula is the reward corresponding to chunk $n$:
\begin{align*}
r_n = q(R_n) - \alpha \lvert q(R_{n-1}) - q(R_n) \rvert - \beta \phi_n,
\end{align*}
where,
\begin{itemize}
\item $q(R_n)$ is the utility function mapping bitrate $R_n$ to a value. Logarithm utility function $q(R_n) = \log(R_n/R^{\min})$ is used as in \cite{bola, pensieve}, where $R^{\min}$ is the bitrate of the minimum quality level.
\item $\lvert q(R_{n-1}) - q(R_n) \rvert$ represents the difference in the quality levels between two consecutive chunks.
\item $\phi_n$ is the rebuffering time in time step $n$. If the download time $d_n$ of chunk $n$ is greater than the amount of video (in seconds) currently in the buffer, $B_n$, then the rebuffering time is $d_n - B_n$.  Otherwise, there is no rebuffering. Hence, rebuffering time associated to chunk $n$ is given by $\phi_n = \max(0, d_n - B_n)$.
\item $\alpha$ and $\beta$ are the coefficients associated with the quality difference and rebuffering time penalties, respectively. We use the values $\alpha=2.6$ and $\beta=1$, the same as \cite{pensieve}.
\end{itemize}

A \textit{state} is a vector of the following components:
\begin{itemize}
\item estimated network throughput for last 06 video chunks in seconds,
\item download time of last 06 video chunks in seconds,
\item next chunk sizes for all quality levels in Mb,
\item current buffer size in seconds,
\item number of remaining chunks,
\item bitrate at which the last chunk was downloaded in Mbps.
\end{itemize}

In the following subsections, we will present FDRLABR algorithms for clients using two groups of DRL algorithms: a value-based algorithm, \textit{i.e.}, DQN, and policy-based algorithms, \textit{i.e.}, A2C and PPO.

\subsection{FDRLABR with DQN}
DQN is a value-based algorithm \cite{dqn}. Q-network with parameter $w$ estimates Q-values $Q(s,a; w)$ from an observed state $s$ for each action $a$. 
An action is selected according to $\epsilon$-greedy policy, in which the agent selects the quality level corresponding to the maximum Q-value with probability $1-\epsilon$ and explores a random quality level with probability $\epsilon$, in the training phase (see \cite{dqn}).
To make the algorithm more stable, DQN uses an experience replay memory to store the transitions \textit{(state $s$, action $a$, reward $r$, next state $s'$)} of previous steps. In addition, a target network parametrized by $w^-$, that periodically clones the parameters of Q-network, is used to estimate the target value $r + \gamma \max_{a} Q (s', a ; {w}^-)$. 
Each training step updates $w$ to minimize the difference between the Q-value $Q (s, a ; w)$ and the target value with a random batch of transitions from the experience replay memory. 

Alg. \ref{alg:dqn} describes the steps of FDRLABR with DQN. The aggregation server collects the weights of Q-networks of the selected clients, averages them, and sends the average weights back to the selected clients of the next global round. 

\begin{algorithm}[h]
\caption{FDRLABR with DQN.}
\begin{algorithmic}[1]
\State Initialize $w_k$ randomly, experience replay memory $\mathcal{D}_k = \emptyset$ for all client $k$.
\For {each global round $t=1, \ldots, T$}
	\State Server uniformly chooses a set $S^{(t)}$ of $K$ clients.
	\State Server sends $w^{(t)}$ to all clients in  $S^{(t)}$.
		\For {each client $k=1$ in  $S^{(t)}$} \Comment {Local updates}
			\State $w_k = w^{(t)}$.
			\For {each iteration $n$ in each of $E$ episodes}
					\State Take action $a_n$ according to $\epsilon$-greedy policy, observe reward $r_n$ and next state ${s}_n'$.
					\State Append transition $({s}_n, a_n, r_n, {s}_n')$ to $\mathcal{D}_k$.
		\State Update $w_k$ by minimizing the loss function 
		\begin{align*}
		\mathbb{E} \big( r + \gamma \max_{a} Q (s', a ; {w}^-_k) - Q( s, a; w_k ) \big)^2
		\end{align*}
		with a random batch from $\mathcal{D}_k$.
		\State If $n \mod C ==0$, ${w}_k^-={w}_k$.
			\EndFor
			\State Send $w_k$ to server.
		\EndFor
	\State Server updates the weights of the global model by averaging the weights received from clients: 
\begin{align*}
w^{(t+1)}=\frac{1}{K}\sum_{k=1}^{K}w_k.
\end{align*}
\EndFor
\end{algorithmic}
\label{alg:dqn}
\end{algorithm}

\subsection{FDRLABR with actor-critic algorithms}
We apply two actor-critic algorithms for clients' ABR, \textit{i.e.}, A2C and PPO.
The first policy-based algorithm, REINFORCE \cite{Sutton}, uses an actor network, parametrized by $\theta$, to estimate the policy $\pi(a|s; \theta)$ directly. 
Actor-critic algorithms use a baseline to reduce the fluctuation of REINFORCE. 
The value function $V(s; \nu)$ is a common baseline, and a critic network, parametrized by $\nu$, is used to estimate this baseline.  
Advantage actor-critic (A2C) or asynchronous advantage actor-critic (A3C) are well-known actor-critic algorithms.
Using $N$-step update, the advantages are given by
\begin{align*}
A_n = R_n - V(s_n; \nu), n=1,\ldots,N.
\end{align*}
where $R_{n} = r_n + \gamma r_{n+1} + \ldots + \gamma^{N -n- 1} r_{N-1} + \gamma^{N - n} V(s_n; \nu)$ (see Alg. \ref{alg:ac}).

Both A2C and A3C use multiple copied environments in parallel to accelerate the training. 
In A3C, the actor corresponding to each environment runs on a separate thread and updates asynchronously \cite{a3c}, whereas A2C uses synchronous updates on only one thread.
We use A2C in the experiments, however, A3C can be implemented similarly.
The objective function in A2C is given by
\begin{align*}
L(\theta) = \sum_{i=1}^m \sum_{n=1}^N \log \pi(a_n|s_n; \theta)A^{(i)}_n,
\end{align*}
where $m$ is the number of parallel environments used in training.

PPO is also an actor-critic algorithm. To prevent the catastrophic drop in the performance of A2C/A3C, PPO constraints the change in policy between two consecutive training steps by introducing a new clipped surrogate objective \cite{ppo}. PPO has shown a reliable performance and is used in many DRL applications.
The surrogate objective of PPO is given by:
\begin{align*}
L(\theta) = \sum_{i=1}^m \sum_{n=1}^N \min\big( r(\theta) {A}^{(i)}_n , \textrm{clip}(r(\theta), 1-\epsilon, 1+\epsilon){A}^{(i)}_n \big),
\end{align*}
where $\epsilon$ is a small constant for clipping and $r(\theta) = \frac{\pi(a|s; \theta)}{\pi(a|s; \theta_{\textrm{old}})}$ in which $\theta_{\textrm{old}}$ is the parameters of the actor network of the previous update. 

In FDRLABR with actor-critic algorithms, the server averages the weights of both actor and critic networks received from the selected clients in each global round. The detailed algorithm is described in Alg. \ref{alg:ac}.

\begin{algorithm}[h]
\caption{FDRLABR with actor-critic algorithms.}
\begin{algorithmic}[1]
\State Initialize parameters $\theta_k, \nu_k$ randomly for all clients $k$.
\For {each global round $t=1,\ldots,T$}
	\State Server uniformly chooses a set $S^{(t)}$ of $K$ clients.
	\State Server sends $w^{(t)} = (\theta^{(t)}, \nu^{(t)})$ to all clients in  $S^{(t)}$.
		\For {each client $k$ in  $S^{(t)}$} \Comment {Local updates}
			\State Update actor and critic networks
			\begin{align*}						
			\theta_k = \theta^{(t)}, \quad
			\nu_k = \nu^{(t)}.
			\end{align*}
			\For {each iteration in each of $E$ episodes}
				\For {environment $i$ in $m$ copied environments at client $k$}
					\State Run policy $\pi(\theta_k$) for $N$ steps.
					\State Compute advantage  ${A}^{(i)}_1,\ldots, {A}^{(i)}_N$.
				\EndFor
				\State Update $\nu_k$ by minimizing 
				\begin{align*}
				\sum_{i=1}^m \sum_{n=1}^N (R_n^{(i)} - V(s_n; \nu_k)^2.
				\end{align*}
				\State Update $\theta_k$ by maximizing the objective $L(\theta_k)$.
			\EndFor
			\State Send $w_k = (\theta_k, \nu_k)$ to server.
		\EndFor
	\State Server updates the weights of global model by averaging the weights $w_k$ received from clients: 
\begin{align*}
w^{(t+1)}=\frac{1}{K}\sum_{k=1}^{K}w_k.
\end{align*}
\EndFor
\end{algorithmic}
\label{alg:ac}
\end{algorithm}

\section{Performance evaluation}
\label{sec:result}
\subsection{Simulation setting}
We use an event-driven simulation that allows a video player to play a 240-second video in less than one second, similar to \cite{pensieve}. The maximum buffer size of the video players is 20 seconds. The video used in the simulation is Big Bug Bunny with seven quality levels 700, 900, 2000, 3000, 5000, 6000, and 8000 Kbps. The video chunk length is four seconds~\cite{video}.
To show the effectiveness of FDRLABRs in different environments, we generate 100 agents with environments different in network bandwidth and round-trip-time.



\textbf{Network bandwidth:}
We use two real-trace datasets: a broadband dataset~\cite{Fcc19} and a 4G LTE Dataset \cite{Raca18}.
The \textbf{broadband dataset} contains over 1 million throughput traces provided by US Federal Communications Commission (FCC). The data are from the ``download speed'' category of the September 2019 collection with 10-second granularity \cite{Fcc19}.
The \textbf{4G dataset} is collected by Irish mobile operators with five mobility patterns: static, pedestrian, car, bus, and train. It contains 135 traces, about 15 minutes per trace at 1-second granularity \cite{Raca18}. 
For each FCC or 4G dataset, we generate 2000 320-second traces, 1000 traces have mean throughputs greater than 2~Mbps, and the other 1000 traces have mean throughputs less than 2~Mbps. 
In each group of traces, we randomly select 80\% for training and the remaining 20\% for testing. 
Each client chooses a random trace in the training set for each episode in the training phase.
In training, each agent randomly selects one of each bandwidth group for training.

We utilize Stable Baseline3 \cite{SB3} library to implement DQN, A2C, and PPO for clients. Stable Baseline3 includes a set of reliable implementations of DRL algorithms and is used in many DRL applications.
We use fully connected neural networks with 64 nodes for each hidden layer. 
We tuned the number of hidden layers for the algorithms.
Table~\ref{table:rlparameter} lists some tuned hyper-parameters for three algorithms. The not-listed hyperparameters are used with the default values provided by Stable Baseline3. 
In FDRLABR with A2C and PPO algorithms, we use $n$-step update with five steps and one environment at each client.

\begin{table}[h!]
	\centering
	\caption{Tuned hyper-parameters of DRL algorithms}
	\begin{tabular}{p{6cm} p{2cm}}
	\toprule
	\textbf{Parameters}     & \textbf{Values}	\\
	\midrule
	\multicolumn{2}{l}{FDRLABR w/DQN tuned parameters:}\\
	\quad q-network &  [64, 64]	\\ 
	\quad activation function   &	Tanh\\
	\quad learning rate	&  0.0005 \\
	\quad batch size	& 128	\\ 
	\quad target network update period ($C$)	& 25	\\
	\quad exploration fraction & 0.5 \\
	\quad exploration final $\epsilon$ & 0.05 \\
	\quad discount factor ($\gamma$)	&0.9 \\	
	\midrule
	\multicolumn{2}{l}{FDRLABR w/A2C tuned parameters:}\\
	\quad actor network &  [64, 64, 64]	\\ 
	\quad critic network &  [64, 64]	\\ 
	\quad activation function   &	Tanh\\
	\quad learning rate	&  0.0005 \\
	\quad discount factor ($\gamma$)	&0.9 \\	
	\midrule
	\multicolumn{2}{l}{FDRLABR w/PPO tuned parameters:}\\
	\quad actor network &  [64, 64, 64]	\\ 
	\quad critic network &  [64, 64, 64]	\\ 
	\quad activation function   &	Tanh\\
	\quad learning rate	&  0.0001 \\
	\quad discount factor ($\gamma$)	&0.9 \\	
	\bottomrule
\end{tabular}
\label{table:rlparameter}
\end{table}

\subsection{Results}
\begin{figure}[h]
	\centering
\includegraphics[width=12cm]{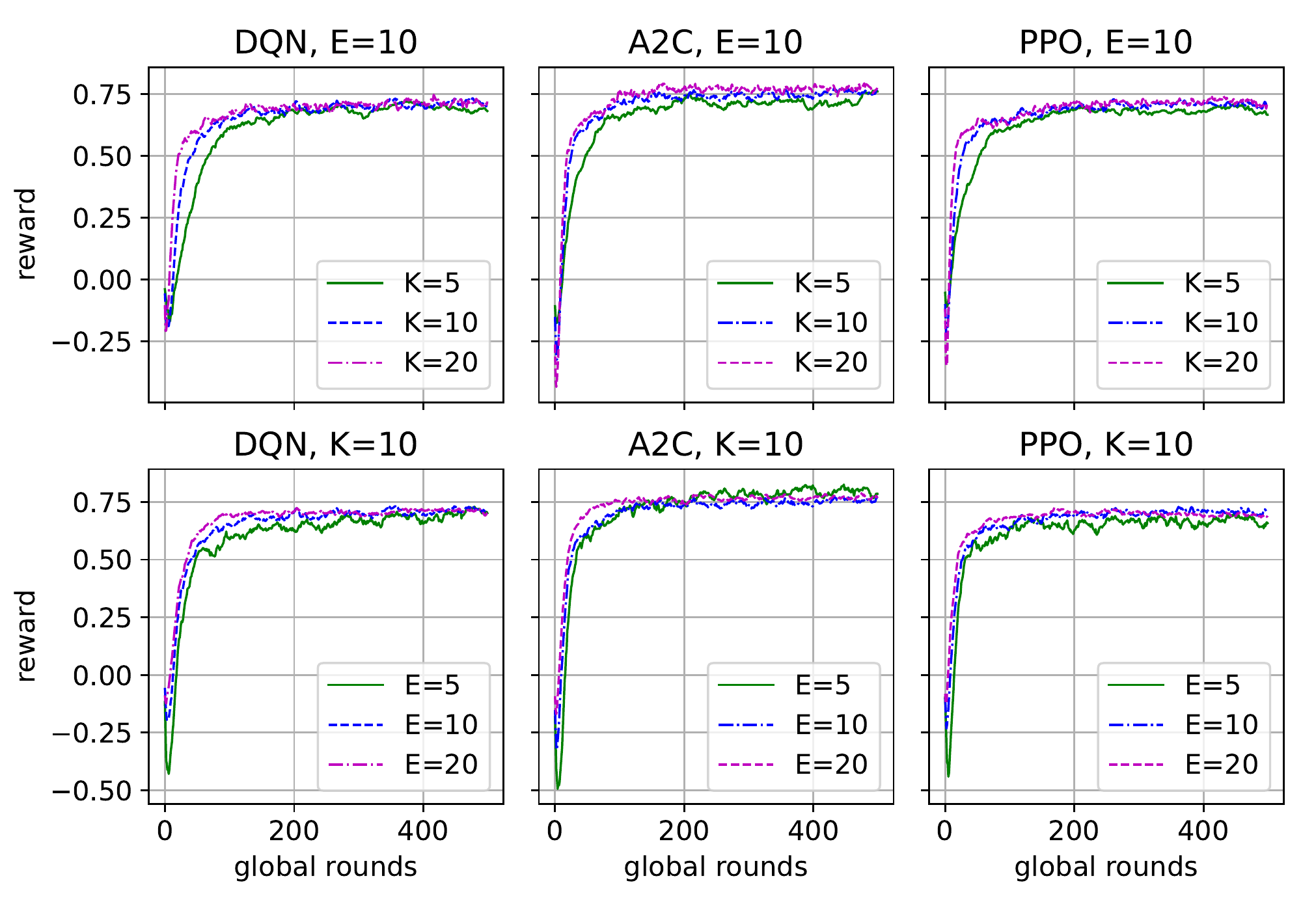}
	\caption{Convergence of FDRLABR with DQN, A2C, and PPO when varying the number of selected users ($E=10$) and varying the number of local episodes ($K=10$).}
	\label{fig:convergence}	
\end{figure}

DRL algorithms are known to be very sensitive to different initial points \cite{drlmatters}.
We train the proposed algorithms with 500 global rounds in five runs.
Fig. \ref{fig:convergence} shows the convergences of the average rewards of 100 clients in five runs in the training phase of FDRLABR with DQN, A2C, and PPO algorithms. 
All the plots show that the more selected users per round ($K$) or the more local episodes the selected users run per round ($E$) yields a better convergence rate and higher rewards. This observation also agrees with the characteristics of FedAvg algorithm \cite{McMahan17}.

We evaluate FDRLABR algorithms with the best models.
Table \ref{table:rewardsbest} compares the test results of FDRLABRs with other ABR methods, \textit{i.e.}, throughput-based (THGHPUT)~\cite{smooth}, buffer-based (BOLA)~\cite{bola}, and CONSTANT (which always chooses 5 Mbps) methods.
The numbers in the table are the average values of 800 episodes corresponding to 800 test traces.
FDRLABR algorithms yield higher average rewards than the other algorithms' rewards, except for FDRLABR with A2C with $K=10$ and $E=10$.
In some cases, FDRLABR with A2C results in lower rewards and much higher variations than FDRLABRs with PPO and DQN. This result shows that FDRLABR with A2C is not as stable as with DQN or PPO, which is also a drawback of A2C.

\begin{table}[h]
	\centering
	\caption{Rewards of ABR methods. FDRLABR's  values are the average rewards and deviations in five runs.}
	\resizebox{0.9\columnwidth}{!}{%
	\begin{tabular}{p{2cm}ccccc}
	\toprule
	\textbf{ABR methods}		&&&\textbf{Rewards}&&\\
\toprule
Constant		&&&-1.427&&\\
\midrule
THGHPUT		&&&0.701	&&\\
\midrule
BOLA			&&&0.742	&&\\
\midrule
FDRLABR&
\shortstack{\textbf{$K=5$}\\ \textbf{$E=10$}} &
\shortstack{\textbf{$K=10$}\\ \textbf{$E=5$}}	&
\shortstack{\textbf{$K=10$}\\ \textbf{$E=10$}} &
\shortstack{\textbf{$K=10$}\\ \textbf{$E=20$}} & 
\shortstack{\textbf{$K=20$}\\ \textbf{$E=10$}} \\
\midrule
\quad w/DQN	& \textbf{0.846$\pm$0.02}& \textbf{0.833$\pm$0.038} & \textbf{0.876$\pm$0.021} &\textbf{0.885$\pm$0.015} &\textbf{0.878$\pm$0.017}	\\
\midrule
\quad w/A2C	& 0.792$\pm$0.127& 0.823$\pm$0.03 & 0.735$\pm$0.125 &0.817$\pm$0.024 &0.83$\pm$0.031 \\
\midrule
\quad w/PPO	& 0.812$\pm$0.034& 0.822$\pm$0.016 & 0.856$\pm$0.023 &0.834$\pm$0.034 &0.855$\pm$0.025 \\
\bottomrule
\end{tabular}
}
\label{table:rewardsbest}
\end{table}

\begin{figure*}[!h]
	\centering
	\includegraphics[width=12cm]{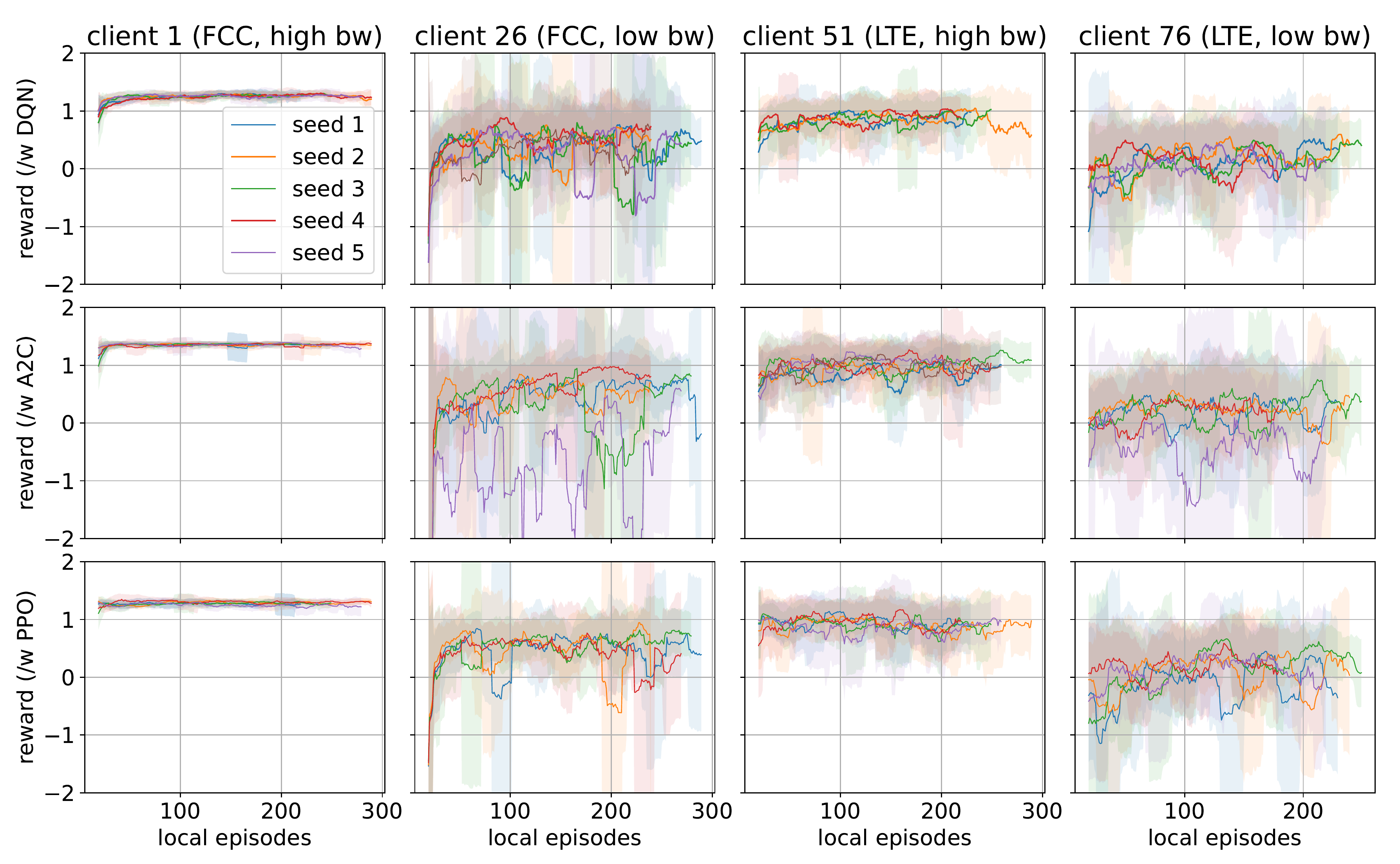}
	\caption{Convergence of FDRLABR with DQN, A2C, and PPO at the users in each group of 5 runs with $K=5, E=10$. The lines are the running average of previous 20 values and the shade regions are the standard deviation.}
	\label{fig:FDRLABR_4grps}	
\end{figure*}

Fig. \ref{fig:FDRLABR_4grps} shows the convergence of FDRLABR with DQN, A2C, and PPO algorithms at clients in each group with $K=5$ and $E=10$.
The rewards resulting from the high-bandwidth FCC and LTE groups have much lower variations than those from the low-bandwidth ones. In each quality level, the actual bitrates of the video chunks are much smaller than the encoding bitrate \cite{video}; indeed, many chunks' bitrates are less than half of their encoding bitrates. Hence, the average reward of each episode in high bandwidth groups is very high. FDRLABR with A2C has the highest variation among the three FDRLABR algorithms, which is also shown in Table~\ref{table:rewardsK5_04grps}.

Table~\ref{table:rewardsK5_04grps} shows the test rewards corresponding to four groups of test traces with $K=5$ and $E=10$. We see that the rewards resulting from the high-bandwidth traces are much higher than those from the low-bandwidth traces.
In most cases, the rewards of FDRLABR algorithms are higher than those of the other algorithms.

   \begin{table}[!htb]
	\centering
	\caption{Average rewards of users in 04 group of environment trained with $K=5$ and $E=10$.}
	\resizebox{0.9\columnwidth}{!}{%
	\begin{tabular}{p{2cm}cccc}
	\toprule
\shortstack{\textbf{ABR}\\ \textbf{methods}}&
\shortstack{\textbf{FCC}\\ \textbf{high bw}}&
\shortstack{\textbf{FCC}\\ \textbf{low bw}}	&
\shortstack{\textbf{LTE}\\ \textbf{high bw}}&
\shortstack{\textbf{LTE}\\ \textbf{low bw}}\\
\toprule
Constant	& 1.123	&-4.244	&0.536	&-3.122\\
\midrule
THGHPUT		&1.239	&0.409	&0.852	&0.305\\
\midrule
BOLA		&1.283	&0.521	&0.995	&0.170\\
\midrule
FDRLABR w/DQN	& 1.333$\pm$0.041 & \textbf{0.687$\pm$0.033} & 1.012$\pm$0.038 & \textbf{0.403$\pm$0.041}\\
\midrule
FDRLABR w/A2C	& 1.357$\pm$0.012	& 0.558$\pm$0.221	& 1.013$\pm$0.048	& 0.239$\pm$0.23\\
\midrule
FDRLABR w/PPO	& \textbf{1.362$\pm$0.005}	& 0.624$\pm$0.054 & \textbf{1.024$\pm$0.01} & 0.239$\pm$0.076\\
\bottomrule
\end{tabular}
}
\label{table:rewardsK5_04grps}
\end{table} 

\section{Conclusions}
\label{sec:con}
We have proposed FDRLABR, a framework combining federated learning with different DRL-based bitrate adaptation algorithms. 
Three DRL algorithms are used at clients, \textit{i.e.}, DQN, A2C, and PPO. 
The clients experience different network environments; however, the trained models yield a better QoE than the traditional methods.


%
%
%
%

\end{document}